\newif\ifsubmode
\submodefalse
 
\newif\ifprintfig
\printfigfalse

\ifsubmode
  \documentclass{aastex}
  \usepackage{epsfig}
  \received{}
  \accepted{}
  \journalid{}{}
  \articleid{}{}
\else
  \documentclass[preprint]{aastex}
  \usepackage{epsfig}
\fi
 
\shortauthors{B\"oker \ea}
\shorttitle{Nuclear Cluster in NGC\,4449}

\newcommand{\ea}{et al.}

\newcommand{\myr}{\>{\rm Myr}}

\newcommand{\mpc}{\>{\rm Mpc}}

\newcommand{\mum}{\>{\mu {\rm m}}}

\newcommand{\msun}{\>{\rm M_{\odot}}}

\newcommand{\as}{ ^{\prime\prime}}
\newcommand{\am}{^{\prime}} 
\newcommand{\bdm}{\begin{displaymath}} 
\newcommand{\edm}{\end{displaymath}}
\newcommand{\beq}{\begin{equation}} 
\newcommand{\eeq}{\end{equation}} 
\newcommand{\bit}{\begin{itemize}} 
\newcommand{\eit}{\end{itemize}} 
\newcommand{\ben}{\begin{enumerate}} 
\newcommand{\een}{\end{enumerate}}
\newcommand{\bfi}{\begin{figure}[htb]} 
\newcommand{\bpfi}{\begin{figure}[p]}

\newcommand{\av}{\rm A_V}

\newcommand{\halpha}{$\rm H\alpha$}
\newcommand{\hbeta}{$\rm H\beta$}

\begin{document}

\title{A Young Stellar Cluster in the Nucleus of NGC\,4449}

\author{Torsten B\"oker\altaffilmark{1}, Roeland P.~van der Marel} 
\affil{Space Telescope Science Institute, 3700 San Martin Drive, 
       Baltimore, MD 21218}
\email{boeker@stsci.edu, marel@stsci.edu}

\author{Lisa Mazzuca\altaffilmark{2}}
\affil{Goddard Space Flight Center, Greenbelt, MD 20771}
\email{lmazzuca@pop500.gsfc.nasa.gov}

\author{Hans-Walter Rix, Gregory Rudnick} 
\affil{Max-Planck-Institut f\"ur Astronomie, K\"onigstuhl 17, 
	D-69117 Heidelberg, Germany}
\email{rix@mpia-hd.mpg.de, rudnick@mpia-hd.mpg.de}

\author{Luis C. Ho} 
\affil{Observatories of the Carnegie Institution of Washington, 
	813 Santa Barbara Street, Pasadena, CA 91101-1292}
\email{lho@ociw.edu}

\author{Joseph C. Shields} 
\affil{Ohio University, Department of Physics and Astronomy, 
	Clippinger Research Laboratories, 251B, Athens, OH 45701-2979}
\email{shields@helios.phy.ohiou.edu}



\altaffiltext{1}{Affiliated with the Astrophysics Division, Space Science 
Department, European Space Agency.} 
\altaffiltext{2}{Current address: Department of Astronomy, University
of Maryland, College Park, MD 20742} 

 
\ifsubmode\else
\clearpage\fi

 
\ifsubmode\else
\baselineskip=14pt
\fi


\begin{abstract}  
We have obtained 1-2$\>$\AA\ resolution optical Echellette spectra of
the nuclear star cluster in the nearby starburst galaxy NGC\,4449.
The light is clearly dominated by a very young ($6-10\myr$) 
population of stars. For our age dating, we have
used recent population synthesis models to interpret
the observed equivalent width of stellar absorption features
such as the H\,I Balmer series and the Ca\,II triplet around 8500$\>$\AA . 
We also compare the observed spectrum of the nuclear cluster 
to synthesized spectra for stellar populations of varying ages.
All these approaches yield a consistent cluster age. Metallicity
estimates based on the relative intensities of various ionization lines 
yield no evidence for significant enrichment in the center of this 
low mass galaxy: the metallicity of the nuclear cluster is about one 
fourth of the solar value, in agreement with independent estimates
for the disk material of NGC\,4449. 
\end{abstract}
\section{Introduction}\label{intro}
Recent observations with the Hubble Space Telescope (HST) at both optical
\citep{phi96,mat99} and near-infrared \citep*[NIR,][]{car98,boe99a} wavelengths
have revealed that a compact, photometrically distinct star cluster is often 
present in the center of spiral galaxies of all Hubble types. 
The formation mechanism of such clusters is puzzling, especially
in late-type spirals which by definition do not possess a prominent bulge. 
In these cases, the gravitational potential is very shallow, and it is
far from obvious how the gas could fall all the way to the center to 
form a dense star cluster. 

In this paper, we describe the results of a pilot program to investigate
the processes of nuclear cluster formation. More specifically, we have
performed 1-2$\>$\AA\ resolution optical Echellette spectroscopy of
the nucleus of NGC\,4449, which is part of a small sample of galaxies 
with prominent nuclear star clusters that were identified from a 
recent NIR survey with HST \citep{boe99a}. 
The morphological classification of NGC\,4449 has been the subject of 
some confusion. It was originally classified as Irr by \citet{hub36} 
and \citet{san61}, but it is listed as Sm in \citet{san81}. As 
\cite{vdb95} points out, the presence of a stellar nucleus might
be a useful criterion to distinguish between the two classes, since
Irregulars usually do not have a well-defined nucleus - in contrast
to late-type spirals, as discussed above.
The extent, morphology, and dynamics of its H\,I gas
suggest that NGC\,4449 has undergone some interaction in the past,
a fact that might well account for its unusually high star formation
activity \citep{hun99}. It also has a prominent stellar bar which 
covers a large fraction of the optically visible galaxy. Both the nuclear
cluster and the stellar bar are evident in Figure~\ref{fig_image}, which
shows a ground-based R-band image by \cite{fre96}.

The fact that the nucleus of NGC\,4449 is
also prominently seen in the far ultraviolet (FUV) 
\citep{hil98} and in various emission lines \citep{sab84,may94,hil98,boe99a} 
indicates that it is the site of ongoing star formation. 
NGC\,4449 is somewhat unusual in this regard, since emission-line
surveys have shown that the nuclei
of late-type galaxies often are devoid of ionized gas \citep[e.g.][]{boe99a}. 
It appears that NGC\,4449 is one of the few late-type
spirals that is ``caught in the act'' of ongoing star formation
in its nucleus. 

NGC\,4449 was chosen for a case study
because of its high cluster-to-disk contrast which allows us to
isolate the light of the nuclear cluster from the
underlying galaxy. From the disk-subtracted cluster 
spectrum, we derive the equivalent width (EW) of a number of stellar 
absorption features. We then use stellar population synthesis methods 
to constrain the star formation history of the nuclear cluster
and to determine the age of its dominant stellar population.
In \S~\ref{sec:data}, 
we describe our observations and the data reduction methods.
Section~\ref{sec:results} describes
the quantitative analysis of a number of emission lines
and stellar absorption features, as well as our spectral
synthesis efforts.
From the results of these various diagnostic tools, we 
deduce an age of about $\rm 6-10\,Myr$ for the
nuclear cluster. In \S~\ref{sec:discussion}, we discuss our results
and compare them to a recent study by \cite*{gel01}
which confirms our age estimate from a complementary dataset. We also
briefly discuss suggested mechanisms of nuclear cluster formation 
and their possible impact on the evolution of the host galaxy.
\section{Observations and data reduction} \label{sec:data}
The data described here were taken in April 1999 at the Steward Observatory 
$90\as$ Bok reflector. We used the Boller \& Chivens spectrograph 
in Echellette mode with a $2\as$-wide slit
to obtain spectra with 1-2$\>$\AA\ resolution
covering the range $\approx 3600-9000\>$\AA . The average seeing over
the observation was $2.5\as$.

Following \cite{wil94}, we used standard IRAF\footnote{IRAF
 is distributed by the National Optical Astronomical Observatories,
 which are operated by AURA, Inc. under contract to the NSF.}
routines in the {\it LONGSLIT} and {\it ECHELLE} packages
to perform the following data reduction steps. After the data frames were 
bias-subtracted and trimmed using the {\it CCDPROC} routine, we corrected 
for bad pixels. We obtained a bad pixel mask by ratioing two domeflat 
exposures with different integration times, and replaced the bad pixels with 
a median filter. A second pass through {\it CCDPROC} then flatfielded the 
data using a median-combined set of domeflat exposures. 

We obtained three 30 minute exposures of NGC\,4449 which were median-combined
after verifying that the brightness profile was properly centered on the
slit for all three exposures. Using the APALL task, we then extracted the
multi-order spectrum of the galaxy over 12 subapertures of $1\as$
each, thus retaining the spatial information. 
Spectra of the standard star Hiltner\,600 were extracted
with the {\it DOECSLIT} task and used to measure the Echelette sensitivity
function, based on spectrophotometry published by \cite{mas88}.
Also within {\it DOECSLIT}, we determined 
the dispersion correction from spectra of a Helium-Argon lamp.
Both flux and wavelength calibration were then applied
to the galaxy spectrum, before we used {\it SCOMBINE} to combine all orders 
into one two-dimensional spectrum covering the full wavelength range.

Finally, we averaged the central $2\as$ into a
one-dimensional spectrum of the nucleus, as well as two stripes 
of $3\as$ width each, located $3.5\as$ on either side of the nucleus. 
Subtracting the two resulting 1-d spectra from each other
eliminates the sky emission lines and isolates the light of the nuclear
star cluster from the underlying galaxy disk. The simple linear
interpolation of the underlying disk is likely to leave
a small residual contribution to the spectrum of the nuclear
cluster. Higher order interpolation clearly is desirable,
but proved unfeasible over the full wavelength range of the
spectrum. However, for limited wavelength ranges, we verified 
that the results of our analysis are robust against such 
variations in the methodology of the background subtraction. We
conclude that the errors introduced by the imperfect interpolation
of the galaxy surface brightness profile are small.

As the result of the described data reduction process, 
the flux- and wavelength-calibrated spectrum between 3600 and 9000$\>$\AA\ 
of the central $2\as$ of NGC\,4449 is shown in Figure~\ref{fig_spectra}.
The red end of the spectrum is subject to some residual fringing with
peak-to-peak amplitudes smaller than 10\% of the continuum. 
This slightly affects the uncertainty of the the continuum level 
for measurement of the Ca triplet absorption lines (see \S~\ref{cat}).
The large residuals near 7580$\>$\AA\ are due to imperfect correction
of telluric oxygen absorption (``A-band''), as are smaller residuals at
6280$\>$\AA\ and 6860$\>$\AA .

\section{Results}\label{sec:results}
\subsection{ Stellar Absorption Features}\label{subsec:ew_analysis}
In order to measure the equivalent width (EW) of various 
stellar absorption features,
we rectified and normalized the spectrum. The line and continuum 
windows that we used for continuum fitting and EW measurement
are defined in \citet*{gon99} for the Balmer and HeI series,
and in \citet*{dia89} for the CaII triplet (CaT) around 8500$\>$\AA . 
They are listed in Table~\ref{tab:absorption}, together with all measured
EWs. Figure~\ref{fig_rectspec} shows the rectified spectrum over two
exemplary regions which contain some higher order Balmer lines and 
the CaT, respectively. The analysis of the Balmer absorption is
complicated by the presence of emission lines. We took the following 
approach to subtract the line emission before measuring the EW of
the absorption features. We iteratively subtracted Gaussian components 
(typically one or two) until the resulting ``pure'' absorption
spectrum of each Balmer line could be well fitted by a 
Voigt-profile. Figure~\ref{fig_method} demonstrates the process.
The sum of the fitted Gauss components then yields the emission line 
fluxes for the Balmer series. Fitting synthesized spectra of 
stellar populations as described in \S~\ref{subsec:modeling} yields an 
alternative approach. After subtraction of the best fitting
model spectrum, the residuals contain only the line emission.
The line fluxes derived with both methods are consistent. 
We conservatively estimate the uncertainty in the derived EWs of
the Balmer absorption to be 10\%. In what follows, 
we compare the measured EWs to evolutionary synthesis models.
\begin{table}
\dummytable \label{tab:absorption}
\end{table}
\subsubsection{Balmer absorption}
In Figure~\ref{fig_balmer}, we compare our measurements to
evolutionary synthesis models of \cite{gon99}. Each panel contains
the time evolution of the EW of a specific absorption feature.
The grey horizontal bar denotes the respective measured EW, the width
of the bar indicates the uncertainty of 10\%. The crossing point
of the bar with the model yields a solution for the cluster age. 
All features yield an age of $6-10\myr$ for the dominant
stellar population in the nuclear cluster of NGC\,4449. This result
is rather independent of the detailed IMF or the star formation
mode (continuous or instantaneous). For all reasonable
models, the rise in the Balmer EW occurs at about the same time, so
that the crossing point of the measurements with the model curve is
robust to within a few million years. A theoretical second solution, 
namely an extremely old cluster with an age around 10 Gyr is ruled out 
from the evidence that follows.
\subsubsection{Calcium triplet}\label{cat}
Because the CaT absorption is produced in cooler and more
evolved stars than the Balmer series, it provides an independent 
age estimator. In Figure~\ref{fig_rectspec}b, we show the spectral region
comprising the CaII triplet around 8500$\>$\AA\ which also contains
the line and continuum windows used for our analysis. These are also
listed in Table~\ref{tab:absorption} and are identical to
those used by \citet*{gar98} to derive theoretical
EWs of the CaT from evolutionary synthesis models. 

We compare the EW sum of the two strongest features ($\lambda$8542 and 
$\lambda$8662) as measured from our data to the \cite{gar98} models 
(their grid I) in Figure~\ref{fig_calcium}. 
The horizontal line again denotes the measured EW of the CaT and
the shaded region its estimated uncertainty which is 
dominated by the uncertainty in the continuum level due to 
residual fringing on the red end of the spectrum. Despite the conservative
error estimate, the very deep CaT absorption in the nuclear cluster of 
NGC\,4449 unambiguously confirms a young cluster age between 5 and $20\myr$.

Figure~\ref{fig_calcium} also seems to indicate a high metallicity of
at least solar value. However, we caution that the low-metallicity
stellar evolutionary tracks used in the population synthesis models
have been shown to systematically underpredict the red supergiant (RSG) 
features \citep{ori99}. This is a longstanding
problem that also manifests itself in the inability of current
models to reproduce the observed blue-to-red supergiant ratio in
galaxies \citep[e.g.][]{lan95}. As a consequence, the predicted CaT EW for
low-metallicity RSG populations falls significantly below the
observed values in, e.g., young clusters in the LMC \citep{bic90}. 
For individual stars, the CaT EW exceeds 9$\>$\AA only
in red supergiants \citep{dia89}. Certainly, any single-age population
of stars that has a CaT EW as high as the nuclear cluster of NGC\,4449 
has to contain red supergiants with ages between 5 and $20\myr$, 
irrespective of its metallicity. 
In summary, Figure~\ref{fig_calcium} confirms a young cluster age, but
does not provide a reliable way to constrain its metallicity. 
In \S~\ref{subsubsec:metallicity}, we will explore alternative methods to 
derive the metallicity based on nebular emission lines.
\subsection{Spectral modeling}\label{subsec:modeling}
As an additional test of the above results we have compared the
spectrum of the nuclear cluster in NGC\,4449 to spectral models, using
the entire available wavelength range. For this we first created model
spectra of simple stellar populations (SSPs), i.e. populations that
form in instantaneous, short-lived starbursts. Models were calculated
for ages on a logarithmically spaced grid with a spacing of $0.5$
dex. We used the 1996 version of the \cite{bru93} models, which assume
a \cite{sal55} IMF between 0.1 and $125\msun$ and solar metallicity. 
The observed spectrum is then modeled as a linear superposition of these
SSPs, which essentially allows an arbitrary star formation history
(approximated as a weighted sum of delta functions). We find the
weights that optimize the fit to the data, allowing also for a 
redshift due to the systemic velocity of the galaxy. 
The $\chi^2$ fit was performed with the software of
\cite{rix92}, developed originally for the analysis of galaxy 
kinematics.

The best fit to the observed spectrum was obtained with a model that
has the large majority of its light ($>90$\%) in the SSP of age $10^7$
years. The data-model comparison is shown in
Figure~\ref{fig_specfit}. The overall shape of the spectrum as well as
the depth of the Balmer absorption features are well
reproduced. However, the Ca\,II feature at 3934$\>$\AA , as well as some
other metal lines, are significantly underestimated in the fit.  This
discrepancy is possibly due to a metallicity mismatch. We used
template spectra with solar abundances for the modeling because we 
did not have lower metallicity tracks with sufficient spectral resolution 
available. However, as we will show in \S~\ref{subsubsec:metallicity}, the
nuclear cluster in NGC\,4449 does actually have subsolar metallicity.
Nonetheless, the
shape of the $4000\>$\AA\ break and the depth of the Balmer series absorption
features depends only slightly on metallicity, so that the quality of the
fit should yield a good age estimate despite the non-perfect choice of 
metallicity.
We conclude that the spectral modeling confirms the results of our previous 
analysis that the nuclear cluster in NGC\,4449 is young.
\subsection{Emission lines}\label{subsec:emission}
As can be seen from Figure~\ref{fig_spectra}, the spectrum of the 
nuclear star cluster of NGC\,4449 -- after subtraction of the 
circum-nuclear light -- contains a number of strong emission lines. 
After subtracting the continuum fit described in \S~\ref{subsec:modeling}, 
we measured the fluxes of all significant emission lines (except
H9 and H10) using {\it SPECFIT} as implemented in IRAF 
\citep{kri94}. The residual spectrum was subdivided into three segments and a
fit with $\chi^2$ optimization was performed, assuming a linear fit
to any residual continuum and Gaussian profiles in a common velocity
frame for the emission lines. The fluxes of H9 and H10
were measured as described in \S~\ref{subsec:ew_analysis}
to more accurately correct for the underlying absorption. The resulting 
fluxes on an absolute scale and relative to H$\beta$ are listed in 
Table~\ref{tab:emission}, along with their 1-$\sigma$ uncertainties.
\begin{table}
\dummytable \label{tab:emission}
\end{table}
\subsubsection{Extinction}\label{subsubsec:extinction}
Extinction of the nebular emission can be quantified from reddening of
the spectrum as indicated by the Balmer emission line ratios. For the
intrinsic relative line strengths, we employed predictions from \cite{hum87}
assuming Case~B recombination in a $10^4$ K gas with
$n_e = 100$ cm$^{-3}$ (e.g., $\rm H\alpha /H\beta$ = 2.86). The electron 
density can be constrained from the ratios of [\ion{O}{2}]
$\lambda\,3729/\lambda\,3726 = 1.38 \pm 0.02$ and [\ion{S}{2}]
$\lambda\,6716/\lambda\,6731 = 1.29 \pm 0.05$, both of which suggest
$n_e \lesssim 100$ cm$^{-3}$ \citep[e.g.][]{ost89}. For the
extinction analysis, we employed the reddening law from \cite*{car89} 
with $R_V=3.1$. The measured Balmer emission
fluxes ratioed to \hbeta\ then imply extinction values of $\rm A_V =
$(1.1, 0.85, 1.4, 1.8, 0.47, 2.6) mag for (\halpha, $\rm H\gamma$,
$\rm H\delta$, $\rm H\epsilon$, H8, H9). The measured flux for H8 is almost
certainly contaminated by emission in \ion{He}{1} $\lambda$3890, which
would account for the low value of $\rm A_V$ suggested by this line.  The
other features show a general trend of increasing $\rm A_V$ with
higher-order Balmer lines, which is probably an indication that the
line fluxes are still slightly underestimated due to stellar
absorption.  We ultimately adopted an extinction value of $\rm A_V = 1.1$
mag based on the \halpha/\hbeta\ ratio, since among the Balmer
lines these features have the highest signal-to-noise ratio and are
expected to suffer the least from underlying absorption. The majority
of this extinction is evidently intrinsic to NGC\,4449, since the
foreground Galactic extinction is only 0.06 mag \citep*{sch98}.
The $\rm A_V$ adopted here is about one magnitude higher
than the value derived by \cite*{ho97} using the same
intrinsic line ratio. On the other hand, \cite{ho97} removed the 
stellar continuum by subtracting generic template spectra in order to 
measure the emission-line strengths. The difference of about
30\% in the measured line ratio needed to reconcile the two measurements
can probably be explained by any one or a combination of the
following reasons: a slight mismatch of the continuum template in the
\cite{ho97} analysis, the fact that they used a somewhat larger aperture, 
or differences in spectral resolution (which affect the ability to separate 
the emission core from the absorption trough).
\subsubsection{Metallicity}\label{subsubsec:metallicity}
The relative line fluxes can be used to obtain an independant estimate
for the metallicity in the nucleus of NGC\,4449.  The ``bright-line''
method originally proposed by Pagel et al. (1979) makes use of the
quantity 
\beq \rm 
R_{23}\equiv (I_{3727}+I_{4959}+I_{5007})/H_{\beta}
\eeq
as a measure of metallicity. For the NGC\,4449 nucleus, the
reddening-corrected line fluxes imply $\rm R_{23} = 8.2 \pm 0.2$. A recent
analysis illustrating the available calibrations of $\rm R_{23}$ is provided
by \cite*{kob99}. Their Figure~8 indicates that our
measured value of $\rm R_{23}$ falls at or near the turnaround point between
the low- and high-metallicity branches for this diagnostic; this
result implies 12 + log(O/H) $\approx 8.3-8.4$, or about one fourth of 
the solar value (12 + log(O/H) = 8.9). This abundance for
the nucleus is in good agreement with the average value for the disk
of NGC\,4449 of 8.32 reported by \cite{lis98}.
Abundances can be inferred directly if diagnostics of the nebular
electron temperature are available. Table~\ref{tab:emission} 
lists a detection of the auroral [\ion{O}{3}] $\lambda$4363 line, which 
in combination with [\ion{O}{3}]$\lambda\lambda$4959, 5007 can be used 
for this purpose. We used the dereddened measurements to calculate the 
O/H abundance using the methods described in \cite{oey00}. The large 
error bar on the [\ion{O}{3}] $\lambda$4363 flux translates into substantial 
uncertainties in the derived quantities, with a resulting temperature 
of $16700 \pm 3500$ K and 12 + log(O/H) = $7.8 \pm 0.3$. The latter 
abundance is consistent at the 2$\sigma$ level with the result based on 
the $\rm R_{23}$ diagnostic.
\subsection{Mass and color of the nuclear cluster}\label{sec:mass}
Assuming a cluster age of $10\myr$, a simple estimate of the cluster 
mass can be obtained from its V-band luminosity. Our flux-calibrated
spectrum after subtraction of the underlying galaxy light
(Figure~\ref{fig_spectra}) yields 
$\rm F_{550\,nm}\,\approx\,3.5\times 10^{-15}\>ergs\,s^{-1}\,cm^{-2}\,\AA^{-1}$,
or $m_V\,=\,15.0$.
This value can be compared to evolutionary synthesis models for
single-age populations such as those in the Starburst99 package
\citep{lei99}. For a Salpeter IMF with
$m_{low}=1\msun$ and $m_{up}=100\msun$, the observed cluster luminosity
requires a mass of about $4\times 10^{5}\msun$ (after correction
for $\rm A_V=1.1$, for a distance\footnote{derived by assuming that 
NGC\,4449 follows the Hubble flow with $\rm H_0=65\>km\,s^{-1}\,Mpc^{-1}$} 
to NGC\,4449 of $3.9\mpc$ and an age of $10\myr$). 
This cluster mass should represent a lower limit for two reasons.
First, the distance to NGC\,4449 is rather uncertain and has been
estimated as high as $5.4\mpc$ \citep{kra79}. In addition, 
the IMF likely has a lower mass cutoff below the assumed $1\msun$. Adding 
low-mass stars does not change the cluster luminosity much, but 
significantly increases the total mass.

As an additional check of our results, we have measured the color of
the nuclear cluster.
The $\rm (V-I)$ color of the spectrum in Figure~\ref{fig_spectra} is 
$\rm (V-I) = 1.1$. Correcting for an extinction of $\av = 1.1$
as derived from the nebular emission lines (\S~\ref{subsubsec:extinction}),
the intrinsic color of the stellar population is $\rm (V-I) = 0.6$.
This value is most likely a lower limit, since at UV- and optical 
wavelengths, the stellar continuum is 
usually less reddened than the nebular emission lines \citep{cal97}.
Comparison with the \cite{lei99} models shows that a color of
$\rm (V-I) \geq 0.6$ is entirely consistent with a $6-10\myr$ old population.

The mass in the nuclear cluster implied by the stellar continuum can
be compared with that inferred from the nebular flux. The
extinction-corrected H$\beta$ flux of $8.9 \times 10^{-14}$ ergs s$^{-1}$
cm$^{-2}$ corresponds to a total H$\beta$ luminosity of $1.6 \times 10^{38}$
ergs s$^{-1}$. For Case B recombination (\S~3.3.1), this luminosity
requires that hydrogen-ionizing photons be produced and absorbed at
a rate of $Q({\rm H}) = 3.4 \times 10^{50}$ s$^{-1}$ within the \ion{H}{2}
region. Comparison to the Starburst99 models shows that this is
entirely consistent with a $10\myr$ old cluster with a mass of
$4\times 10^{5}\msun$ and a metallicity of about one fourth of the
solar value.

The dereddened \ion{He}{1} $\lambda$5876/H$\beta$ ratio of $0.10 \pm
0.01$ can be compared with the Case B prediction \citep{ben99} 
of 0.14(${{{\rm He}^+/{\rm H}^+}\over {0.1}}$), where 
$\rm He^+/H^+$ is the ionic abundance ratio. The fact that the
observed line ratio is close to the prediction
implies that most of the helium within the nuclear \ion{H}{2} region
is singly ionized. This condition sets a strong limit on the age of
the cluster, since relatively hot stars are required to maintain the
helium ionization. The production rate of helium-ionizing photons
implied by the extinction-corrected \ion{He}{1} $\lambda$5876 flux 
is $2.6 \times10^{49}$ s$^{-1}$. For the above cluster parameters, this 
luminosity is consistent with the Starburst99 prediction at an age of 
$\sim 7\myr$.
\section{Discussion and outlook} \label{sec:discussion}
The results of this study imply that NGC\,4449 can be added to a 
growing list of late-type spiral galaxies that host a compact, young 
star cluster in their nucleus. While accurate spectroscopy - and hence 
a reliable age estimate - of nuclear clusters is available for only a 
few objects such as the Milky Way \citep{kra95}, NGC\,7793 \citep{shi92},
and IC\,342 \citep*{boe99b}, photometric measurements show that many 
clusters are fairly blue, e.g. those in M\,31, M\,33, NGC\,4242, 
or ESO\,359-029 \citep{lau98,mat99}. This indicates the existence of a 
young ($\leq 100\myr$) stellar population in many of these nuclei.

From an HST survey of intermediate-type spirals, 
\cite{car98} find that virtually all galaxies with
exponential bulges (i.e. bulges with an exponential surface brightness 
profile, as opposed to ``power-law'' or ``$R^{1/4}$'' bulges)
host compact nuclear clusters and that their 
luminosity correlates with that of the host galaxy. 
If one assumes that the mass-to-light ratio of the host galaxies 
as a whole is roughly the same for all objects,
the most straightforward interpretation of this result is that more 
massive galaxies have more massive nuclear clusters. On the other hand, 
stellar populations fade with time \citep[e.g.][]{bru93}. Another
possible interpretation therefore is that more massive galaxies host
younger clusters. This intrinsic degeneracy makes age dating of
nuclear clusters from photometric data alone impossible. Spectroscopic
analysis is needed to investigate the age distribution of nuclear
clusters and to decide whether their formation is a one time event
or not. If nuclear clusters indeed form recurrently, the formation
processes likely impact the morphological evolution of the host galaxy.

Of particular interest in this context is a model suggested by 
\citet{fri93} and developed further by \citet*{nor96}. These authors
point out that build-up of a central mass concentration - e.g.
a nuclear star cluster - can dissolve
stellar bars, and lead to the formation of a bulge via collective
bending instabilities \citep{rah91,mer94}. 
The \cite{nor96} simulations show that only 5\% of the combined
disk and bar mass in a central concentration is sufficient to destroy a 
stellar bar on short timescales, leading to a bulge-like distribution 
of stars. In this scenario, repeated cycles 
of bar formation --- gas infall --- cluster formation
--- bar disruption will build the galaxy bulge over time.
If true, the Hubble classification scheme could be naturally explained
as an evolutionary sequence from late to early Hubble types. 

In the case of NGC\,4449, the cluster mass of (at least) 
$4\times 10^{5}\msun$ derived in \S~\ref{sec:mass} would be sufficient
to destroy a bar with a mass up to $10^{7}\msun$. The unusually 
prominent stellar bar in NGC\,4449, which covers a large part of the 
optically visible galaxy, is probably much more massive than this value.
This can be seen from the total luminosity of NGC\,4449, which is listed
in the RC3 \citep{dev91} as $\rm V^0_T = 9.53$, which corresponds to 
$\rm L_V \approx 2\times 10^9\>L_{\odot ,V}$.
Since the optical image is dominated by the stellar bar, and typical
mass-to-light ratios are of order unity, the bar mass should be 
$\approx 10^9\msun$. The fact that it co-exists with the nuclear cluster  
is therefore not inconsistent with the \cite{nor96} scenario.

However, as pointed out before, NGC\,4449 is a special case because
of its unusually high overall star formation rate and its complex 
history with evidence for past interactions. The complex gas flows
produced by a past merger might well explain why -- different from
most other late-type spirals \citep[e.g.][]{boe99a} -- the 
nucleus of NGC\,4449 contains 
large amount of ionized gas. Furthermore, there is no indication of an 
underlying older population of stars in the nuclear cluster of NGC\,4449,
as is the case, e.g., in the Milky Way. 
We point out, however, that for external galaxies where individual
stars cannot be resolved, it is difficult to detect an old stellar 
population underlying a young starburst because the latter is dominating
the luminosity (assuming similar masses for both bursts). It is somewhat 
unclear at this point whether the processes that govern the star formation 
in the nucleus of NGC\,4449 are similar to those in ``typical'' late-type
spirals, or whether NGC\,4449 is an entirely different ``special'' case.

After the present paper had been finished, another study of the nuclear
cluster in NGC\,4449 appeared in the literature \citep{gel01}.
Using ground-based spectroscopy of the $\rm ^{12}CO(2,0)$ and 
$\rm ^{12}CO(3,1)$ absorption features around $2.3\mum$ and UVI 
colors from HST imaging, these authors derived a cluster age 
between 8 and $15\myr$, in good agreement with our estimate of $6-10\myr$. 
Assuming a somewhat lower extinction to the NGC\,4449 nucleus than derived
by us from the Balmer decrement (\S~\ref{subsubsec:extinction}), 
Gelatt \ea\ conclude that the color
of the nuclear cluster is better matched by evolutionary tracks with
a metallicity of Z=0.008, somewhat higher than the average over the
NGC\,4449 disk. However, increasing the extinction by one magnitude 
(roughly the difference between our extinction and theirs) would easily
move the location of the nuclear cluster in their Figure~3 within the range
of the lower metallicity tracks. After accounting for the different
extinction estimates, the derived V-I color of the
nuclear cluster differs by about 0.2 magnitudes in the two studies. 
This could be the result of the somewhat different aperture sizes used
to extract the cluster light ($1.4\as$ for Gelatt \ea\ compared to
$2.0\as$ for our study). Either way, both papers come to very similar
conclusions on the age of the nuclear cluster in NGC 4449, with
complementary methods.

This study is part of a larger observational program that contains
three independent, but related approaches. First, we will use
the WFPC2 onboard the Hubble Space Telescope during Cycle 10 to measure
the structural properties of all nuclear clusters found in a large 
sample of nearby, face-on spiral galaxies of late Hubble type. We will then 
measure the stellar velocity dispersion in the brightest nuclei found
in order to derive the cluster mass, as recently demonstrated 
for the example of IC\,342 \citep{boe99b}.
Finally, we hope to obtain optical spectra with HST/STIS
to derive the stellar population and to constrain the formation 
history of individual nuclear clusters in much the same way as described 
in this paper. We aim to study a large
number of nuclear clusters to allow a statistical analysis of their
properties and dependence on the Hubble type of the host galaxy. Since
ground-based observations are possible only for the brightest nuclear
clusters, they are likely biased towards the young end of their age
distribution. The goal of our study is to eliminate this bias. Measuring
the true age distribution of nuclear star clusters will allow us to
answer the question whether repeated nuclear starbursts are indeed
common in late-type spirals, and, if so, what their duty cycle is.
\section{Summary}
We have presented a new, high-quality spectrum of the nuclear star cluster
of NGC\,4449 which covers the region between 3700 and 9000$\>$\AA .
Analysis of a variety of stellar absorption features leads to the
conclusion that the nucleus of NGC\,4449 has undergone a short-lived
starburst about $6-10\myr$ ago. We do not find evidence for significant
metal enrichment due to multiple previous 
periods of star formation in the nucleus of NGC\,4449.
\acknowledgements
We thank the anonymous referee for useful comments that helped to improve
the presentation of this paper.

\newpage
\figcaption[figure1.ps]{Ground-based R-band image of NGC\,4449 from 
\cite{fre96}. The field is $7\am \times 7\am$, with North up and East to 
the left. The greyscale has been optimized to the dynamic range of the galaxy.
The nuclear cluster is clearly visible.
 \label{fig_image}}

\figcaption[figure2.ps]{Top: spectrum of the nuclear star cluster of NGC\,4449,
taken over a $2\as \times 2\as$ aperture. The underlying disk/bulge 
emission has been subtracted. Bottom: the same spectrum on
an expanded y-scale. The residuals near 7580$\>$\AA ,
6280$\>$\AA , and 6860$\>$\AA\ are due to imperfect correction
of telluric oxygen absorption (``A-band'').
A few sky lines have not been perfectly
corrected, and appear as negative ``spikes''.
 \label{fig_spectra}}

\figcaption[figure3.ps]{Normalized and rectified spectrum of the Balmer
series (top) and the CaT absorption features (bottom). The horizontal bars
denote the line and continuum windows that were used for
measurement of the EW (Table~\ref{tab:absorption}).
 \label{fig_rectspec}}

\figcaption[figure4.ps]{Spectrum of H$\delta$ before (top) and after (bottom)
removal of the line emission. The Gauss and Voigt profile fits that were
used to subtract the emission line and to judge the completeness of 
the line removal are overplotted in the respective panels (dashed lines). 
 \label{fig_method}}

\figcaption[figure5.ps]{Comparison of the Balmer EW with evolutionary
synthesis models of \cite{gon99} for an instantaneous
burst with a \cite{sal55} IMF between $\rm M_{low} = 1\msun$ and 
$\rm M_{up} = 80\msun$. The horizontal grey bar indicates the measured
EW for the nuclear cluster in NGC\,4449, its width corresponds to the 
measurement uncertainty. The crossing point of bar and model indicates 
an age of $\leq 10\myr$.
 \label{fig_balmer}}

\figcaption[figure6.ps]{Comparison of the CaT EW with evolutionary
synthesis models of \cite{gar98} for an instantaneous burst with a 
\cite{sal55} IMF between $\rm M_{low} = 0.8\msun$ and $\rm M_{up} = 100\msun$. 
The horizontal grey bar indicates the measured
EW for the nuclear cluster in NGC\,4449, its width corresponds to the 
measurement uncertainty. The crossing point of bar and model indicates 
an age of $\approx 10\myr$. The low-metallicity models are
highly uncertain, the fact that they do not reproduce the observed
EW should not be considered as evidence for a high
cluster metallicity (see discussion in text).
 \label{fig_calcium}}

\figcaption[figure7.ps]{Results of the spectral modeling.
The modeled spectrum for a $10\myr$ stellar population with a \cite{sal55}
IMF, based on the \cite{bru93} evolutionary tracks (thick line) is 
plotted over the observed spectrum of the nuclear cluster in NGC\,4449.
The spectrum has a resolution of $\approx 1\>$\AA , about 5 times higher 
than the model.
\label{fig_specfit}}

\newpage
\begin{deluxetable}{lccc}
\tablenum{1}
\tablewidth{0pt}
\tablecaption{Absorption features}
\tablehead{
\colhead{Species} & \colhead{Line window} & EW [\AA] & \colhead{Cont. window}
}        
\startdata
H10 $\lambda$3799 & 3783-3813 & 2.81 & a \\ 
H9 $\lambda$3835 & 3823-3853 & 3.72 & a \\ 
H8 $\lambda$3889 & 3874-3904 & 3.96 & a \\ 
H$\epsilon$ $\lambda$3970 & 3955-3985 & 2.01 & a \\ 
H$\delta$ $\lambda$4102 & 4070-4130 & 5.03 & b \\ 
H$\gamma$ $\lambda$4340 & 4311-4371 & 3.42 & c \\
H$\beta$ $\lambda$4862 & 4830-4890 & 3.88 & d \\ 
Ca\,II $\lambda$8498 & 8483-8513 & 4.05 & e \\ 
Ca\,II $\lambda$8542 & 8527-8557 & 6.78 & e \\ 
Ca\,II $\lambda$8662 & 8647-8677 & 4.45 & e \\ 
\enddata
\tablecomments{Continuum windows are defined as follows:\\
(a) 3740:3743,3760:3762,3782:3785,3811:3812,3909:3915\\
(b) 4019:4020,4037:4038,4060:4061,4138:4140,4148:4150\\
(c) 4301:4305,4310:4312,4316:4318,4377:4381,4392:4394,4397:4398\\
(d) 4820:4830,4890:4900\\
(e) 8447-8462,8842-8857 \\
Windows (a)-(d) are taken from \cite{gon99}, (e) is from \cite{dia89} }
\end{deluxetable}
\newpage
\begin{deluxetable}{lccc}
\tablenum{2}
\tablewidth{0pt}
\tablecaption{Emission Line Fluxes}
\tablehead{
\colhead{Species} & \colhead{F [$\rm 10^{-15}ergs/s/cm^2$]} & 
\colhead{F/F(\hbeta)} & \colhead{$\rm F_c/F_c$(\hbeta)} 
}         
\startdata
$[$\ion{O}{2}$]$ $\lambda$3726  & 42.3$\pm$0.6 & 1.56$\pm$0.03 & 2.27$\pm$0.05 \\ 
$[$\ion{O}{2}$]$ $\lambda$3729  & 58.4$\pm$0.6 & 2.15$\pm$0.04 & 3.13$\pm$0.06 \\ 
H10 $\lambda$3798 		& $\leq$0.2 & $\leq$0.01 & - \\ 	
H9 $\lambda$3836 		& 0.9$\pm$0.2 & 0.03$\pm$0.01 & 0.04$\pm$0.02 \\ 	
$[$\ion{Ne}{3}$]$ $\lambda$3869	& 4.2$\pm$0.5  & 0.15$\pm$0.02 & 0.22$\pm$0.03 \\
H8 $\lambda$3889 		& 2.5$\pm$0.4 & 0.09$\pm$0.01 & 0.13$\pm$0.02 \\ 	
H$\epsilon$ $\lambda$3970 	& 2.6$\pm$0.3  & 0.1$\pm$0.01 & 0.13$\pm$0.02 \\ 	 
H$\delta$ $\lambda$4102 	& 5.0$\pm$0.4 & 0.18$\pm$0.02 & 0.24$\pm$0.02  \\ 	
H$\gamma$ $\lambda$4340         & 11.0$\pm$0.3 & 0.41$\pm$0.01 & 0.48$\pm$0.01 \\ 	
$[$\ion{O}{3}$]$ $\lambda$4363  & 1.2$\pm$0.4 & 0.04$\pm$0.01 & 0.05$\pm$0.02 \\ 
H$\beta$ $\lambda$4862	 	& 27.1$\pm$0.4 & 1.00$\pm$0.02 & 1.00$\pm$0.02 \\ 	
$[$\ion{O}{3}$]$ $\lambda$4959 	& 20.0$\pm$0.4 & 0.74$\pm$0.02 & 0.72$\pm$0.02 \\ 
$[$\ion{O}{3}$]$ $\lambda$5007 	& 58.9$\pm$0.4 & 2.17$\pm$0.04 & 2.04$\pm$0.03 \\ 
$[$\ion{He}{1}$]$ $\lambda$5876 & 3.3$\pm$0.2 & 0.12$\pm$0.01 & 0.10$\pm$0.01 \\ 
$[$\ion{O}{1}$]$ $\lambda$6300  & 2.9$\pm$0.4 & 0.11$\pm$0.01 & 0.08$\pm$0.01 \\ 
$[$\ion{N}{2}$]$ $\lambda$6548  & 4.2$\pm$0.4 & 0.15$\pm$0.01 & 0.11$\pm$0.01 \\ 
H$\alpha$ $\lambda$6563         & 109.5$\pm$0.3 & 4.04$\pm$0.06 & 2.82$\pm$0.04 \\ 	
$[$\ion{N}{2}$]$ $\lambda$6583  & 13.5$\pm$0.3 & 0.5$\pm$0.01 & 0.35$\pm$0.01  \\ 
$[$\ion{S}{2}$]$ $\lambda$6716  & 13.0$\pm$0.3 & 0.48$\pm$0.01 & 0.32$\pm$0.01  \\ 
$[$\ion{S}{2}$]$ $\lambda$6731  & 10.1$\pm$0.3 & 0.37$\pm$0.01 & 0.25$\pm$0.01  \\ 
$[$\ion{Ar}{3}$]$ $\lambda$7136 & 3.1$\pm$0.3 & 0.11$\pm$0.01 & 0.7$\pm$0.01 \\ 
\enddata
\tablecomments{Column~2: Line flux as measured from Figure~1. Column~3:
measured line ratio relative to \hbeta . Column~4: line ratio relative 
to \hbeta\ after correction for an extinction of $\rm A_V = 1.1$ mag.
Uncertainties listed are 1$\sigma$.}
\end{deluxetable}
\newpage
\begin{figure}
\vspace{16.0cm}
\end{figure}

\newpage
\begin{figure}
\epsfxsize=18.0truecm
\centerline{\epsfbox{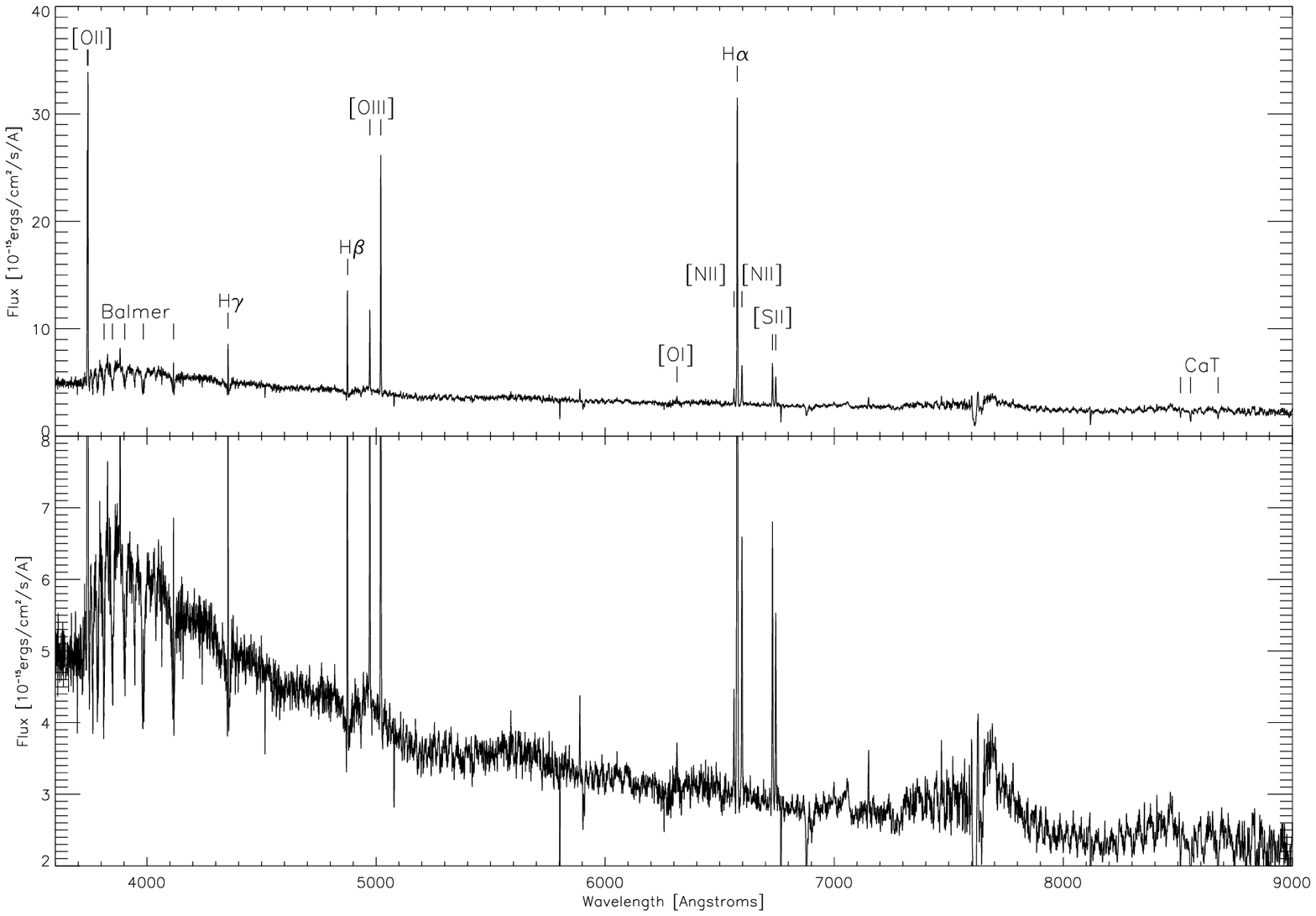}}
\end{figure}

\newpage
\begin{figure}
\epsfxsize=16.0truecm
\centerline{\epsfbox{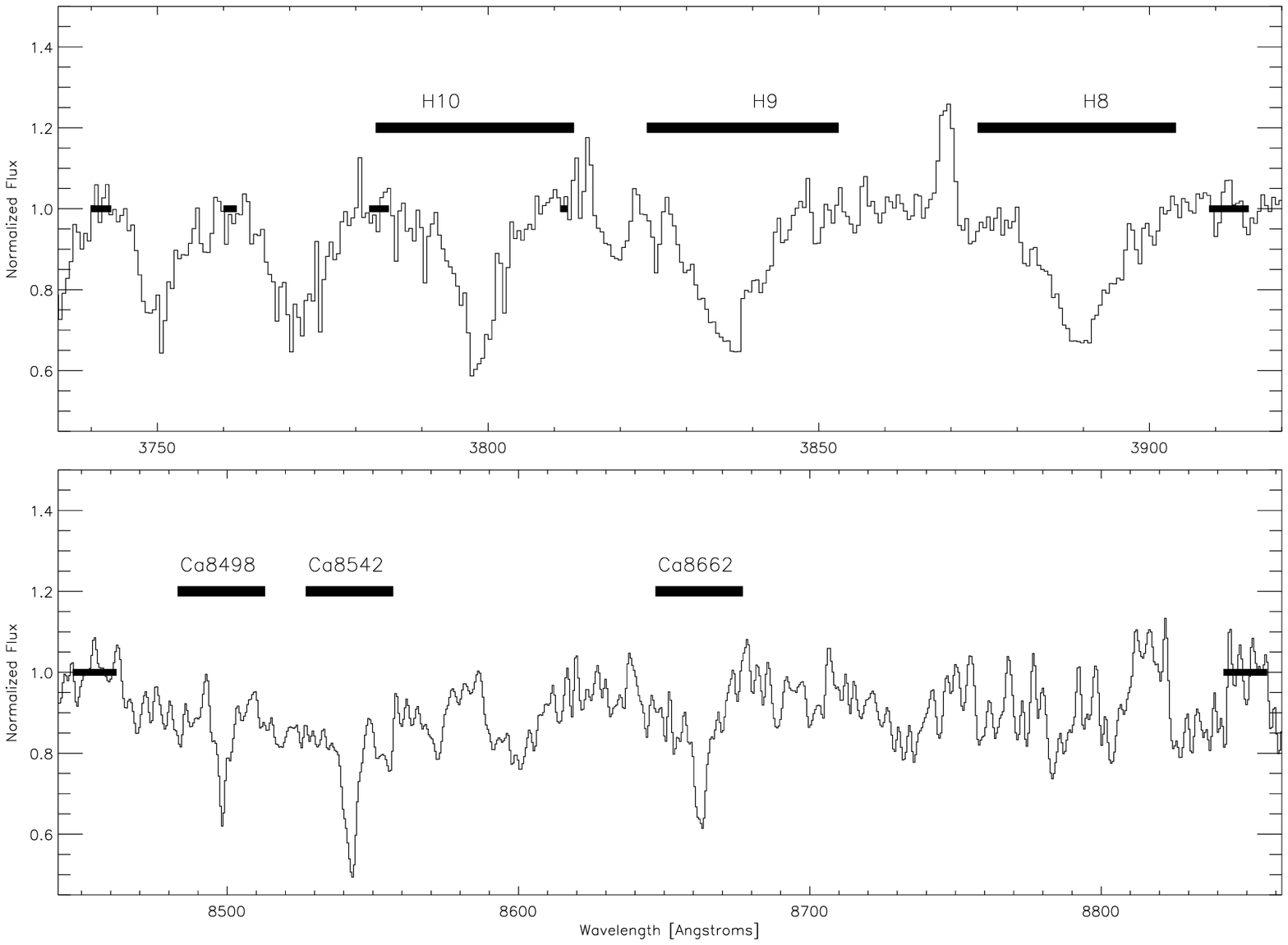}}
\end{figure}

\newpage
\begin{figure}
\epsfxsize=16.0truecm
\centerline{\epsfbox{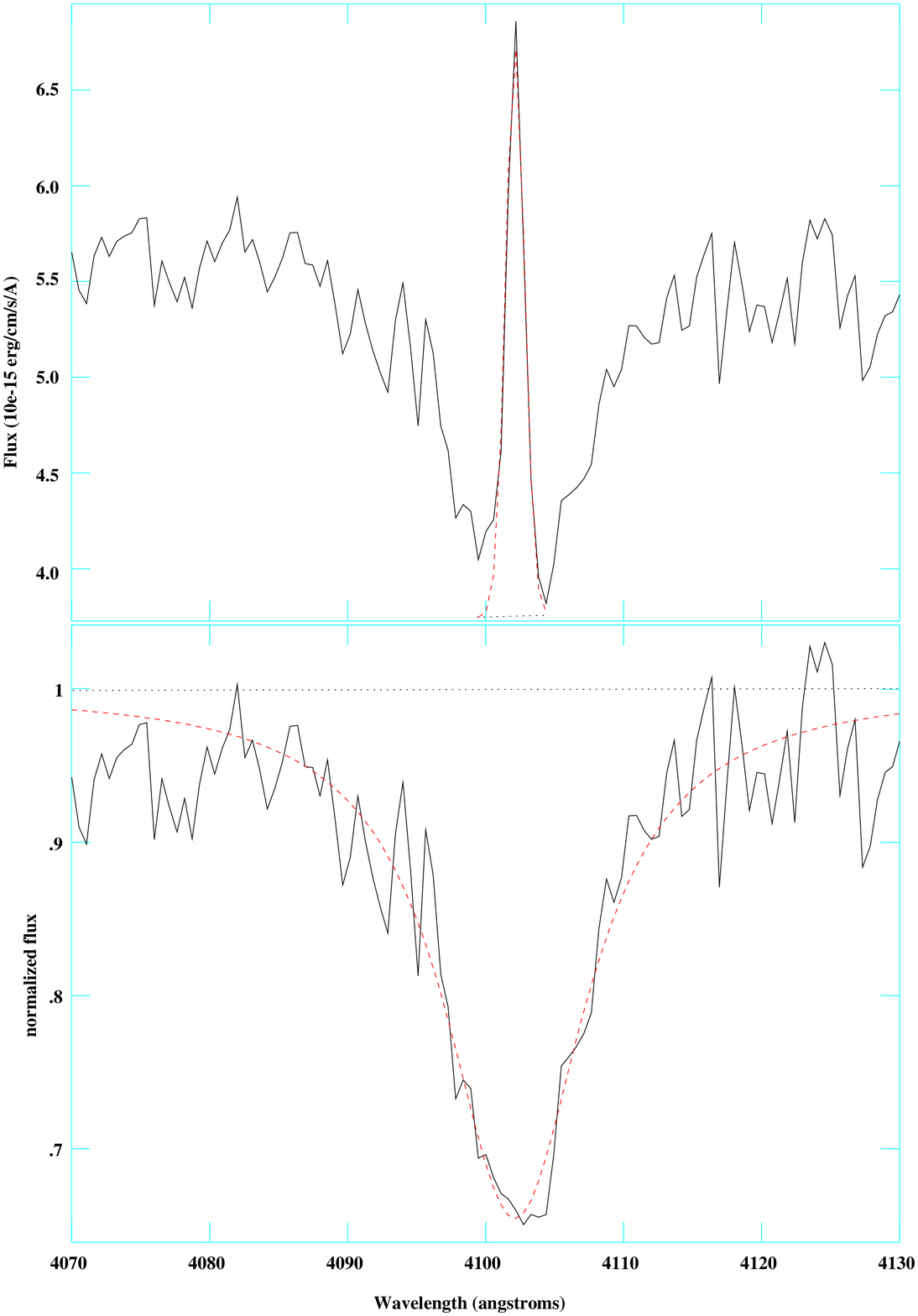}}
\end{figure}

\newpage
\begin{figure}
\epsfxsize=18.0truecm
\centerline{\epsfbox{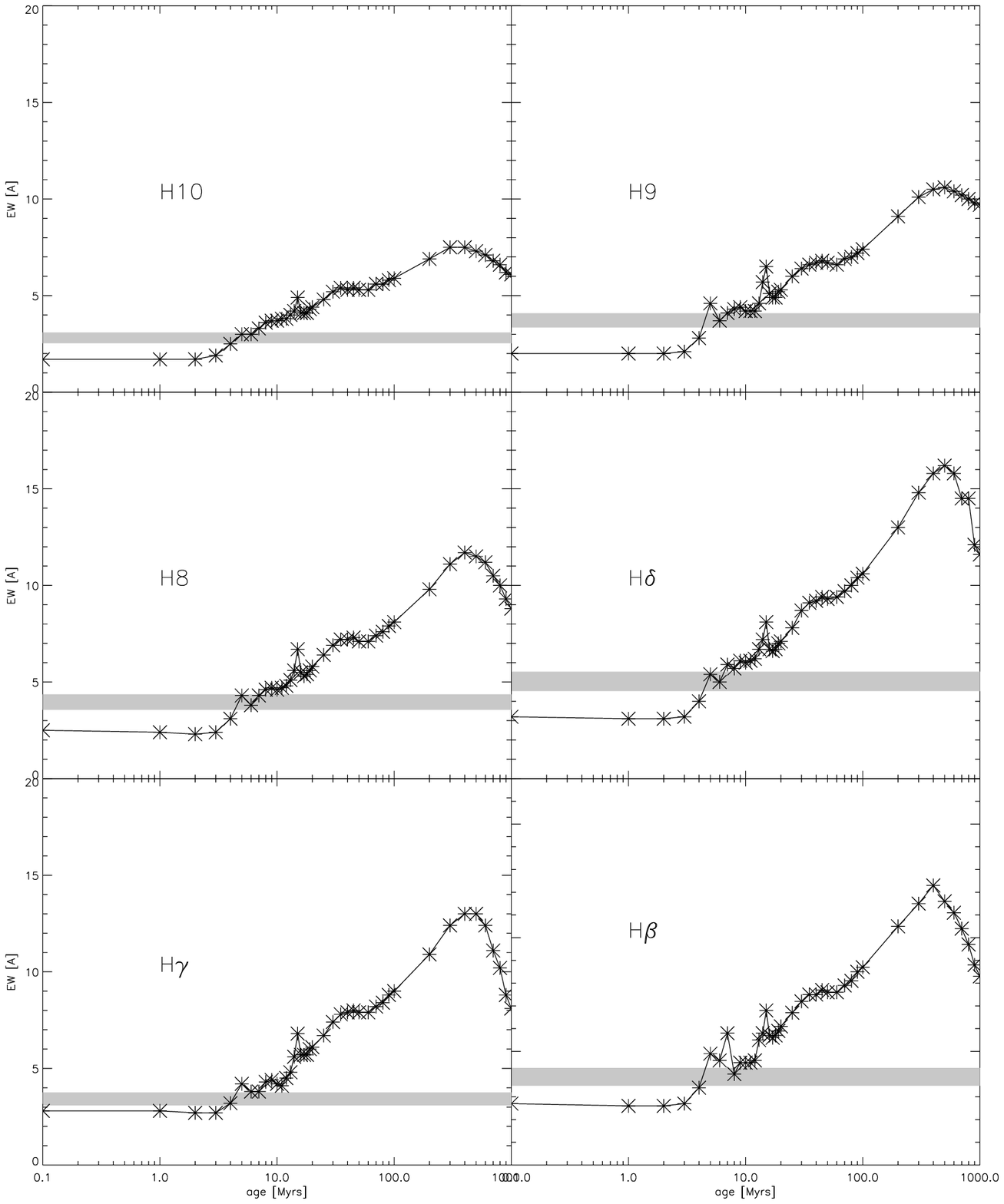}}
\end{figure}

\newpage
\begin{figure}
\epsfxsize=16.0truecm
\centerline{\epsfbox{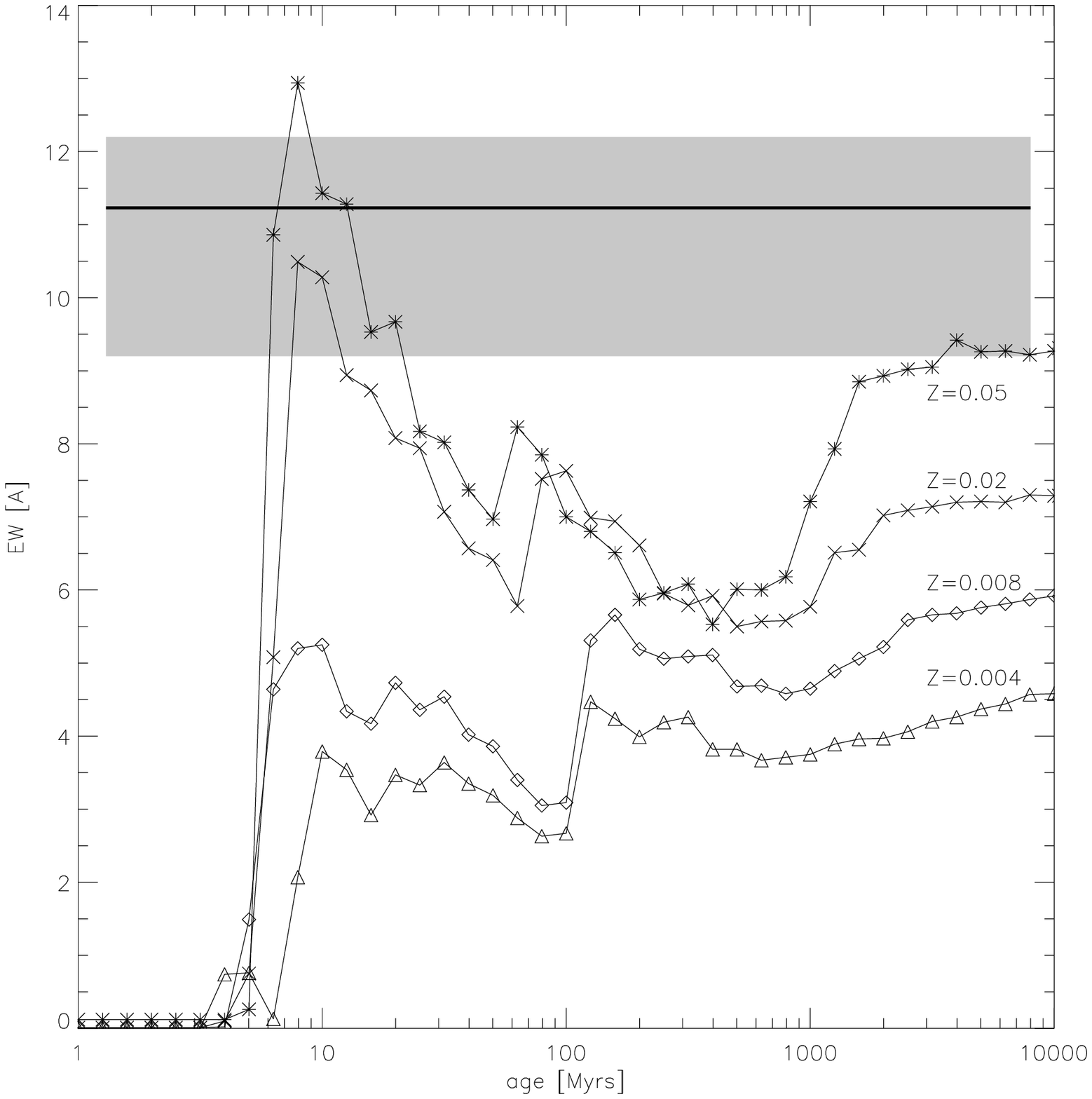}}
\end{figure}

\newpage
\begin{figure}
\epsfxsize=16.0truecm
\centerline{\epsfbox{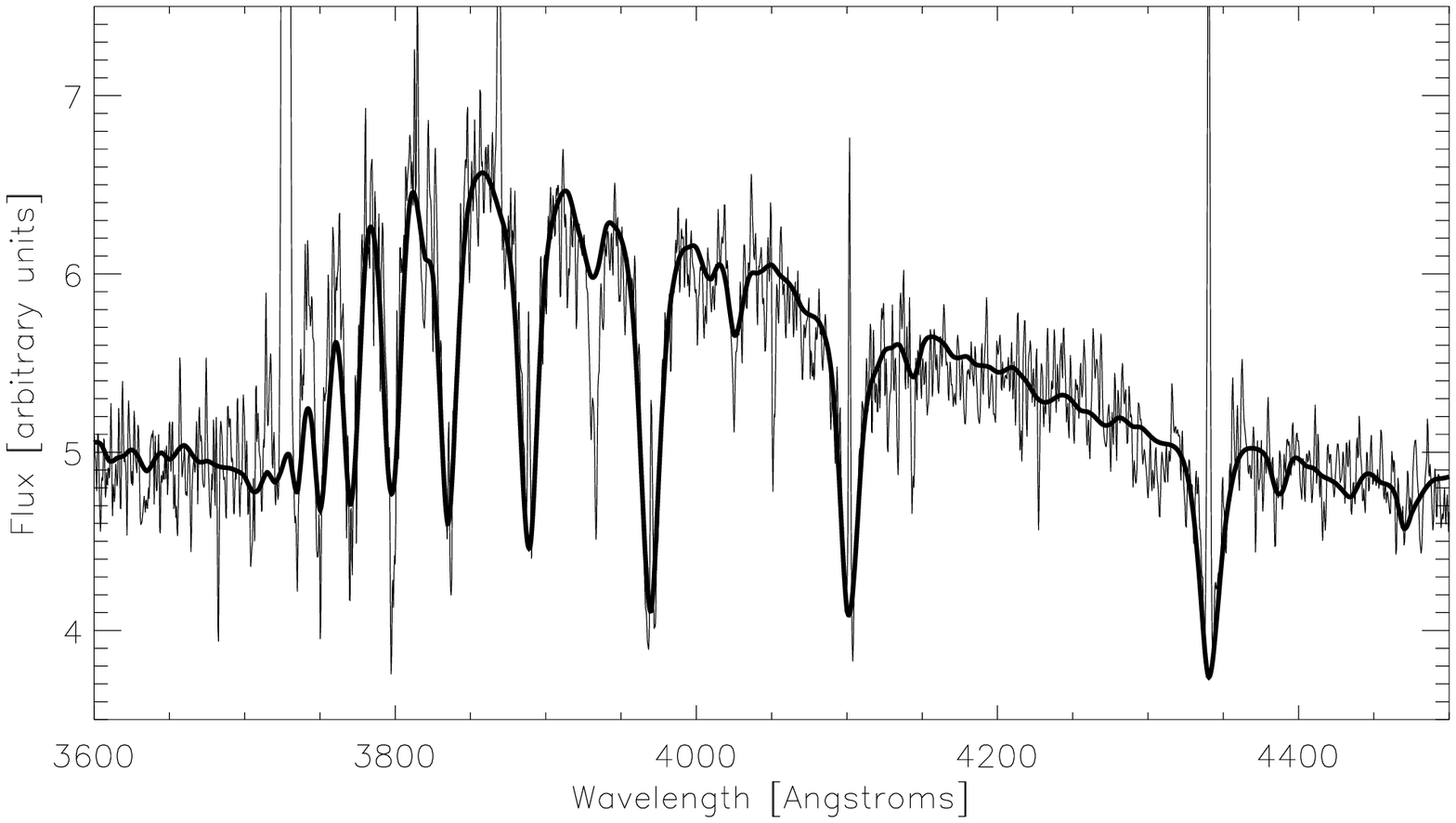}}
\end{figure}


\newpage
\begin{thebibliography}{}
\bibitem[Athanassoula(1992a)]{ath92a}
	Athanassoula, E. 1992a, MNRAS, 259, 328 
\bibitem[Athanassoula(1992b)]{ath92b}
	Athanassoula, E. 1992b, MNRAS, 259, 345
\bibitem[Benjamin \ea (1999)Benjamin, Skillman, \& Smits]{ben99}
	Benjamin, R. A., Skillman, E. D., \& Smits, D. P. 1999, ApJ, 514, 307
\bibitem[Bica, Santos, \& Alloin(1990)]{bic90}
	Bica, E., Santos, J. F. C., Jr., \& Alloin, D. 1990, A\&A, 235, 103
\bibitem[B\"oker \ea (1999a)]{boe99a}
	B\"oker, T., \ea\ 1999a, ApJS, 124, 95 
\bibitem[B\"oker \ea (1999b)B\"oker, van der Marel \& Vacca]{boe99b}
	B\"oker, T., van der Marel, R. P., \& Vacca, W. D. 1999b, AJ, 118, 831
\bibitem[Bruzual \& Charlot(1993)]{bru93}
	Bruzual, A. \& Charlot, S. 1993, ApJ, 405, 538
\bibitem[Calzetti(1997)]{cal97}
	Calzetti, D. 1997, AJ, 113, 162
\bibitem[Cardelli \ea (1989)Cardelli, Clayton \& Mathis]{car89}
	Cardelli, J. A., Clayton, G. C., \& Mathis, J. S. 1989, ApJ, 345, 245  
\bibitem[Carollo \ea (1998)Carollo, Stiavelli, \& Mack]{car98}
	Carollo, C. M., Stiavelli, M., \& Mack, J. 1998, AJ, 116, 68  
\bibitem[Courteau, de Jong \& Broeils(1996)]{cou96}
	Courteau, S., de Jong, R., \& Broeils, A. 1996, ApJL, 457, L73 
\bibitem[de Vaucouleurs(1948)]{dev48}
	de Vaucouleurs, G. 1948, Ann. d'Astrophysique, 11, 247 
\bibitem[de Vaucouleurs \ea (1991)]{dev91}
	de Vaucouleurs, G., de Vaucouleurs, A., Corwin, H., 
	Buta, R. J., Paturel, G., \& Fouque, P. 1991, 
	{\it Third Reference Catalogue of Bright Galaxies}, 
	New York:Springer-Verlag
\bibitem[D{\'\i}az \ea (1989)D{\'\i}az, Terlevich \& Terlevich]{dia89}
	D{\'\i}az, A. I., Terlevich, E., \& Terlevich, R. 1989, MNRAS, 239, 325 
\bibitem[Frei \ea (1996)]{fre96}
	Frei, Z., Guhathakurta, P., Gunn, J. E., \& Tyson, J. A. 1996, 
	AJ, 111, 174  
\bibitem[Friedli \& Benz(1993)]{fri93}
	Friedli, D. \& Benz, W. 1993, A\&A, 268, 65  
\bibitem[Garc{\'\i}a-Vargas \ea (1998)Garc{\'\i}a-Vargas, Moll\'a \& Bressan]{gar98}
	Garc{\'\i}a-Vargas, M. L., Moll\'a, M., \& Bressan, A. 1998, 
	A\&AS, 130, 513
\bibitem[Gelatt \ea (2001)Gelatt, Hunter, \& Gallagher]{gel01}
	Gelatt, A. E., Hunter, D. A., \& Gallagher, J. S. 2001, 
	PASP, in press (February issue)
\bibitem[Gonz\'alez Delgado \ea (1999)Gonz\'alez Delgado, Leitherer \& Heckman]{gon99}
	Gonz\'alez Delgado, R. M., Leitherer, C., \& Heckman, T. M. 
	1999, ApJS, 125, 489
\bibitem[Hill \ea (1998)]{hil98}
	Hill, R. S. \ea\ 1998, ApJ, 507, 179 
\bibitem[Ho \ea (1997)Ho, Filippenko \& Sargent]{ho97}
	Ho, L. C., Filippenko, A. V., \& Sargent, W. L. W. 1997, ApJS, 112, 315  
\bibitem[Hubble(1936)]{hub36}
	Hubble, E. 1936, {\it The Realm of the Nebula}, Yale University Press,
	New Haven
\bibitem[Hummer \& Storey(1987)]{hum87}
	Hummer, D. G., \& Storey, P. J. 1987, MNRAS, 224, 801
\bibitem[Hunter, van Woerden \& Gallagher(1999)]{hun99}
	Hunter, D. A., van Woerden, H., \& Gallagher, J. S. 1999, AJ, 118, 2184  
\bibitem[Kobulnicky \ea (1999)Kobulnicky, Kennicutt, \& Pizagno]{kob99}
	Kobulnicky, H. A., Kennicutt, R. C. Jr., \& Pizagno, J. L. 1999,
	ApJ, 514, 544
\bibitem[Kormendy \& Bender(1996)]{kor96}
	Kormendy. J. \& Bender, R. 1996, ApJ, 464, 119
\bibitem[Kraan-Korteweg \& Tamman(1979)]{kra79}
	Kraan-Korteweg, R. C. \& Tamman, G. A. 1979, Astron. Nachr. 300, 181
\bibitem[Krabbe \ea (1995)]{kra95}
	Krabbe, A. \ea\ 1995, ApJ, 447, L95
\bibitem[Kriss(1994)]{kri94}
	Kriss, G. 1994, in ASP Conf. Ser. 61, Astronomical Data Analaysis
	Software and Systems III, ed. D. R. Crabtree, R. J. Hanisch, \&
	J. Barnes (San Francisco: ASP), 437
\bibitem[Lauer \ea (1998)]{lau98}
	Lauer, T.~R., Faber, S. M., Ajhar, E. A., Grillmair, C. J., \&
	Scowen, P. A. 1998, AJ, 116, 2263
\bibitem[Langer \& Maeder (1995)]{lan95}
	Langer, N. \& Maeder, A. 1995, A\&A 295, 685
\bibitem[Leitherer \ea (1999)]{lei99}
	Leitherer, C., et al. 1999, ApJS, 123, 3
\bibitem[Lisenfeld \& Ferrara(1998)]{lis98}
	Lisenfeld, U. \& Ferrara, A. 1998, ApJ, 496, 145
\bibitem[Massey \ea (1988)]{mas88}
	Massey, P., Strobel, K., Barnes, J. V., \& Anderson, E. 1988, 
	ApJ, 328, 315
\bibitem[Matthews \ea (1999)]{mat99}
	Matthews, L. D. \ea\ 1999, AJ, 118, 208  
\bibitem[Mayya(1994)]{may94}
	Mayya, Y. D. 1994, AJ, 108, 1276  
\bibitem[Merritt \& Sellwood(1994)]{mer94}
	Merritt, D. \& Sellwood, J. A. 1994, ApJ, 425, 551  
\bibitem[Norman \ea (1996)Norman, Sellwood \& Hasan]{nor96} 
	Norman, C. A., Sellwood, J. A., Hasan, H. 1996, ApJ, 462, 114 
\bibitem[Oey \& Shields(2000)]{oey00}
	Oey, M. S., \& Shields, J. C. 2000, ApJ, 539, 687
\bibitem[Origlia \ea (1999)]{ori99}
	Origlia, L., Goldader, J. D., Leitherer, C., Schaerer, D., \&
	Oliva, E. 1999, ApJ, 514, 96
\bibitem[Osterbrock(1989)]{ost89} 
	Osterbrock, D. E. 1989, {\it Astrophysics of Gaseous Nebulae and
	Active Galactic Nuclei}, University Science Books:Mint Valley, CA 
\bibitem[Pagel \ea (1979)]{pag79}
	Pagel, B. E. J., Edmunds, M. G., Blackwell, D. E., Chun, M. S., \&
	Smith, G. 1979, MNRAS, 189, 95  
\bibitem[Phillips \ea (1996)]{phi96}
	Phillips, A.C., Illingworth, G. D., MacKenty, J. W., 
        \& Franx, M. 1996, AJ, 111, 1566  
\bibitem[Raha \ea (1991)]{rah91}
	Raha, A., Sellwood, J.A., James, R., \& Kahn, F.D. 1991, Nat, 352, 411
\bibitem[Rix \& White(1992)]{rix92}
	Rix, H.-W. \& White, S. D. M. 1992, MNRAS, 254, 389
\bibitem[Sabbadin \ea(1984)Sabbadin, Ortolani \& Bianchini]{sab84} 
	Sabbadin, F., Ortolani, S., \& Bianchini, A. 1984, A\&A, 131, 1
\bibitem[Salpeter(1955)]{sal55} 
	Salpeter, E. E. 1955, ApJ, 121, 161
\bibitem[Sandage(1961)]{san61} 
	Sandage, A. 1961, {\it The Hubble Atlas of Galaxies}, Carnegie
	Institution, Washington 
\bibitem[Sandage \& Tamman(1981)]{san81} 
	Sandage, A. \& Tamman, G. A. 1981, {\it A Revised Shapley-Ames 
	Catalog of Bright Galaxies}, Carnegie Institution, Washington 
\bibitem[Schlegel \ea (1998)Schlegel, Finkbeiner, \& Davis]{sch98} 
	Schlegel, D. J., Finkbeiner, D. P., \& Davis, M. 1998, ApJ, 500, 525
\bibitem[Shields \& Filippenko(1992)]{shi92} 
	Shields, J. C. \& Filippenko, A. V. 1992, ASP Conf. series, 31, 267
\bibitem[van den Bergh(1995)]{vdb95}
	van den Bergh, S. 1995, AJ, 110, 613
\bibitem[Willmarth \& Barnes(1994)]{wil94}
      Willmarth, D. \& Barnes, J. 1994, {\it A User's Guide to Reducing 
      Echelle Spectra with IRAF}, National Optical Astronomy Observatories, 
      Arizona
\end{thebibliography}
\end{document}